\documentclass[aps,footinbib,showpacs,twocolumn]{revtex4}

\usepackage{natbib}
\usepackage{amsmath}
\usepackage{amssymb}
\usepackage{graphicx}
\usepackage{dcolumn}
\usepackage{bm}
\usepackage{epsfig}
\begin{document}

\title{Nematic order of model goethite nanorods in a magnetic field}
\author{H.H. Wensink and G. J. Vroege}
\email{g.j.vroege@chem.uu.nl}
\affiliation{Van 't Hoff Laboratory for Physical and Colloid Chemistry, Debye Institute, Utrecht University, Padualaan 8,
3584 CH Utrecht, The Netherlands}
\date{\today}

\begin{abstract}
We explore the nematic order of model goethite nanorods in an external magnetic field within Onsager-Parsons density 
functional theory. The goethite rods are represented by monodisperse, charged spherocylinders with a permanent magnetic 
moment along the rod main axis, forcing the particles to align parallel to the magnetic field at low field strength.
The intrinsic diamagnetic susceptibility anisometry of the rods is negative which leads to a preferred 
perpendicular orientation at higher field strength.  It is shown that these counteracting effects may give rise to intricate 
phase behavior, including a pronounced stability of  biaxial nematic order and the presence of reentrant phase transitions and demixing phenomena. The effect of the applied field on the nematic-to-smectic transition will also be addressed.

\end{abstract}

\maketitle

\section{Introduction}

The effect of external fields on the phase behavior of anisometric colloids (rods and plates) has received considerable attention over the last years, in particular in the domain of density functional theory \cite{evansDFT}. The presence of an interface or wall breaks the translational and orientational symmetry and leads to local structures drastically different from those found in the bulk \cite{roijdijkstraevansEURO,alleninterface,shundyakinterface,harnaudietrich}. Macroscopically different behavior may be brought about for instance by a gravitational field \cite{baulinkhokhlov,baulin}.  If the  buoyant mass of the colloids is sufficiently high, inhomogeneous density profiles  are built up along the vertical dimension of the system. In some cases, this may lead to the formation of multiple phase equilibria  not encountered at zero-field \cite{wensinksedi, schmidtfloatliq,savenkosedi}.  

In addition, much  effort has been put into investigating the behavior of anisometric colloids in an applied electric or magnetic field.  Owing to the fact that the electric polarizability (or magnetic susceptibility) is different along the short and long  axis of the particle, electric or magnetic dipole moments are induced which give rise to an additional energetic contribution to the free energy. The resulting competition between minimal self-energy  and maximal configurational entropy of the rods drastically changes the orientational structure of the system and leads to qualitatively different phase behaviour. 
An important difference with the previously mentioned class of fields is that an external electric or magnetic field is only coupled with the orientational degrees of freedom of the rods or plates and therefore does not directly lead to spatial inhomogeneities. 

 The first systematic attempt to incorporate the effect of these {\em directional} external fields into  the classic Onsager theory \cite{onsager} for lyotropic anisometric particles has been reported by Khokhlov and Semenov \cite{khokhlovsemenov}.  Within a simplified variational approach,  general phase behavior scenarios were presented for both hard and semi-flexible rod systems subjected to various types of directional fields. Later on, similar studies have been carried out within a numerical treatment of the Onsager theory \cite{vargaszalaiNN,vargakronome} and Lee-Parsons \cite{vargasjacksonszalai} and MWDA-type \cite{grafloewen} density functional approaches. In the latter case various (spatially) inhomogeneous liquid crystal states which may occur at high packing fractions, e.g. smectic and (plastic) solids,  were also considered.

Although the experimental study of electro-magnetic field effects on colloidal suspensions has been pioneered a long time ago (see Ref. \cite{lemairefaraday} and references therein), the topic has been subject of renewed interest because of recent experiments on colloidal goethite ($\alpha$-FeOOH)
suspensions \cite{davidsongoethiet,lemairefaraday}. These systems consist of charged, bar-shaped nanorods with peculiar  magnetic properties. These particles not only possess  a permanent magnetic moment directed along their longitudinal axis, originating from uncompensated surface spins
 within the anti-ferromagnetic crystal lattice, but also an enhanced magnetic susceptibility along their short axes. This means that  additional magnetic moments are induced  {\em perpendicular} to the main axis upon applying an external field. These unique properties become manifest in particular in concentrated, nematic suspensions  subjected to magnetic fields below 1 Tesla. At low field strengths, the induced moments are weak and the permanent ones give rise to  enhanced  nematic alignment along the field direction. However, at high field strengths the induced moments are dominant and cause the rods to orient with their longitudinal axes perpendicular to the field. The associated reorientation of the nematic director can be clearly observed from X-ray scattering measurements \cite{lemaireSAXS}.
 
Although  the realignment phenomenon can be  understood directly from the counteracting effects of the permanent and induced magnetic dipoles, very little is known about the overall phase diagram of goethite systems as a function of applied field strength. In Ref. \cite{davidsongoethiet}, a first attempt has been made towards a global understanding of the phase behavior but the analysis there was restricted to ideal systems, described by a simple Boltzmann distribution for the rod orientations, and weakly correlated systems described by an expansion of the Onsager free energy up to first order in the degree of nematic order. Both approaches may be used for very dilute isotropic suspension where particle interactions are not too important, but lack predictive power for dense systems where interactions lead to strong deviations from the ideal Boltzmann orientation distribution associated with the magnetic energy.
 
In this paper, we give a full numerical analysis of the phase behavior of model goethite suspensions,  
starting from  Onsager-Parsons density functional theory. The numerical approach not only allows us to 
 explore the full nematic density range, it also provides a reliable way to probe subtle changes in the nematic orientation symmetry as a function of the applied field strength, in particular transitions between uniaxial and biaxial nematic states.
 We stress that the model goethite system we consider in this paper is a strongly simplified one. The particles are considered as monodisperse, charged spherocylinders interacting via electrostatic repulsions. Weak attractive van der Waals forces are present in the experimental systems \cite{lemairethesis}, but are difficult to incorporate theoretically and are not considered here. The interaction energy between the total dipole moments on the rods is estimated to be of order $10^{-5} k_{B}T$ \cite{lemairethesis} and can therefore be safely neglected. Moreover, the particles' considerable size polydispersity (along all three particle axes) not only leads to a wide variety of particle shapes, but also a strong concomitant spread in the magnetic and electrostatic properties (e.g. surface charge). Therefore, all quantities presented here pertaining to the electrostatic and magnetic properties  should be considered as {\em typical} values rather than quantitative averages.

This paper is constructed as follows. In Sec. II the Onsager-Parsons approach will be presented and adapted for charged particles and the presence of a directional field. In Sec. III we quantify the average magnetic and electrostatic properties of the goethite rods. 
 Depending on the relative contribution of the (average) permanent and induced dipole moments, several phase diagram scenarios for goethite were constructed. They will be discussed in Sec. V. In the next section, we scrutinize the implications of the magnetic field on the nematic-to-smectic transition by means of a first-order bifurcation analysis.
 In the Appendix we supplement our numerical work  with an analytic variational approach based upon the Gaussian approximation. This provides us with a simple tractable theory for strongly ordered nematic states.

 \section{Onsager-Parsons theory}
 
The starting point of our analysis is the magnetic energy of a single goethite rod which consists of two parts;  a contribution for the remanent magnetic moment along the main rod axis,
linear in the magnetic field strength $B$ and  one representing the induced  magnetisation perpendicular to the main axis
which depends quadratically on $B$. Following Ref. \cite{davidsongoethiet}, the total magnetic energy can be written as 
\begin{equation}
\beta U_{m}(\cos \theta) = -J B {\cal P}_{1}(\cos \theta) + K B^{2} {\cal P}_{2}(\cos \theta) \label{magnenergy}
\end{equation}
in terms of the Legendre polynomials ${\cal P}_n$ with $\theta$ the angle between the main rod axis and the direction of the magnetic field. The quantities $J$ and $K$, with dimensions  $T ^{-1}$ and $T^{-2}$ ($T$ is Tesla), respectively, are related to the (average) remanent dipole moment $\mu_{r}$ and the diamagnetic susceptibility anisometry $\Delta \chi=\chi_{\parallel}-\chi_{\perp}<0$ of the nanorods via:
\begin{equation}
J=\beta \mu_{r}, \hspace{1cm} K = \beta \Delta \chi v_{0}/3 \mu_{0} \label{jeeka} 
\end{equation}
with $v_{0}$ the rod volume and $\mu_{0}$ the magnetic permeability in vacuum. All quantities will be specified in Sec. III-A.

As a first approximation, the bar-shaped goethite rods are modelled as (uniaxial) spherocylinders with equal length $L$ and diameter 
$D$, bearing a uniform electric surface charge. Following Onsager \cite{onsager}, the charged rods interact via an 
effective hard core repulsion, characterized by an effective diameter $D_{\text{eff}}>D$ which depends  on the 
charge density on the particle and the ionic strength of the solvent.  The Onsager-Parsons free energy of the system in the presence of an external directional field can be cast into the following functional form \cite{onsager,stroobantslading}:
\begin{eqnarray}
\frac{\beta F}{N} & \sim& \ln c  + \sigma [f] + c g_{\text{P}}(\phi) \left \{ \rho[f] + h \eta[f] \right \}  \nonumber \\
&& + \left \langle \beta U_{m}(\cos \theta) \right \rangle _{f} \label{free}
\end{eqnarray}
where the brackets denote an orientational average according to some singlet orientation distribution function (ODF) $f(\Omega)$ characterizing the average orientational configuration of the system in terms of the solid angle $\Omega$. 
Here, $c$ and $\phi$ denote the {\em effective} (dimensionless) number density $c=(\pi/4) NL^{2}D_{\text{eff}}/V$ and 
packing fraction $\phi = Nv_{\text{eff}}/V$ with $v_{\text{eff}}=(\pi/4)LD^{2}_{\text{eff}}+(\pi/6)D^{3}_{\text{eff}}$ the effective volume of the spherocylinder.  The last term in Eq. (\ref{free}) represents the external magnetic contribution [cf. Eq. (\ref{magnenergy})], whereas $\sigma$ and $\rho$  quantify  the orientational and packing entropy, respectively, 
defined by the following angular averages:
 \begin{eqnarray}
  \sigma [f] &=& \left \langle \ln 4 \pi f(\Omega) \right \rangle _{f} \nonumber \\
\rho [f] &=& \frac{4}{\pi} \left \{ \left \langle \left \langle \sin \gamma (\Omega,\Omega ') \right \rangle \right \rangle _{f} +  \frac{\pi D_{\text{eff}}}{L} + \frac{2 \pi}{3} \left ( \frac{D_{\text{eff}}}{L} \right )^{2} \right  \}  
 \hspace{0.5cm}  \label{rhosigma} 
\end{eqnarray}
where $\gamma$ is the angle between two spherocylinders with orientations $\Omega$ and $\Omega^{\prime}$.
The second term in $\rho[f]$ arises from end-cap contributions to the (electrostatic) repulsion between two short spherocylinders
and is strictly speaking only valid in the isotropic state \cite{odijkendeffects}. 
The contribution $\eta$ expresses the so-called `twisting effect' arising from
the orientation-dependent nature of the electrostatic interaction \cite{onsager,brenner}:
\begin{eqnarray}
\eta [f] &=& \frac{4}{\pi} \left \langle \left \langle -\sin \gamma (\Omega,\Omega ') \ln  [ \sin \gamma (\Omega,\Omega ') ] \right \rangle \right \rangle _{f} \nonumber \\
&& - \left ( \ln 2 - 1/2 \right ) \rho[f]  \label{eta}
\end{eqnarray}
The importance of this effect is quantified by a twisting parameter $h =\kappa^{-1}/D_{\text{eff}}$, defined as the ratio of 
the Debye screening length $\kappa ^{-1}$ and the effective rod diameter \cite{stroobantslading}.  All properties pertaining to the electrostatic interactions will be specified in Sec. III-B for the case of goethite. The function $g_{P}(\phi)$ originates from the Lee-Parsons \cite{Parsons,Lee87,Lee89}
rescaling of the original second virial theory and is proportional to the Carnahan-Starling excess free energy ${\cal F}_{CS}^{ex}$ (in units $k_{B}T$ per particle) for a hard sphere fluid \cite{carnahanstarling} via:
\begin{equation}
g_{P}(\phi) = \frac{{\cal F}_{CS}^{ex}}{4\phi} = \frac{1-(3/4) \phi}{(1-\phi ) ^2}
\end{equation}
The equilibrium ODF is determined by applying a formal minimization of the free energy. This yields the following self-consistency condition:
\begin{eqnarray}
f(\Omega ) &=& Z \exp\left [-\frac{8}{\pi} c g_{P}(\phi )  \int \omega (\Omega, \Omega ' ) f(\Omega ') d \Omega ' \right ]  \nonumber \\
&& \times \exp \left [ - \beta U_{\text{m}} (\cos \theta ) \right ] \label{iter} 
\end{eqnarray}
where $Z$ is obtained from the normalization condition $\int f(\Omega)d\Omega =1$.  $\omega$ is the orientation-dependent part of the second virial coefficient for two charged spherocylinders 
at fixed solid angles $\Omega$ and $\Omega '$ \cite{stroobantslading}:
\begin{eqnarray}
\omega (\Omega, \Omega ' ) &=& \sin \gamma (\Omega,\Omega ') \nonumber \\
   && \times  \left \{ 1 - h  \left ( \ln [ \sin \gamma (\Omega,\Omega ')] + \ln 2 - 1/2 \right)  \right \}
   \hspace{0.4cm}
\end{eqnarray}
The solid  angle $\Omega$ is conveniently parametrized in terms of a polar angle $ 0 \leq \theta \leq \pi $ and an azimuthal one $ 0 \leq  \varphi \leq 2 \pi$, so that $d\Omega=\sin\theta d \varphi$.
Throughout the remainder  of this text,  $ \theta $ always refers to the angle between the rod main axis and the direction of the magnetic field. The azimuthal angle $ \varphi $ then describes the projection of the rod axis onto the plane {\em perpendicular} to the field. Eq. (\ref{iter}) is solved iteratively according to a discretization scheme outlined in Ref. \cite{Herzfeldgrid} using a 2D grid of angles $\{ \theta_{i}, \varphi_{j} \}_{i,j=1,N}$  of mesh size $N>30$. 

To specify the orientational symmetry in the various nematic states we define the following nematic order parameters: 
\begin{eqnarray}
S_{n} &=& \left < {\cal P}_{n}(\cos \theta)\right >_{f}, \hspace{1cm} n=1,2 \nonumber \\
\Delta &=& \left < \sin ^{2} \theta \cos 2 \varphi    \right >_{f} 
\end{eqnarray}
The first one, $S_{1}$, quantifies the degree of dipolar order due to the permanent dipoles at finite field strengths.
Note that for non-dipolar rods  $S_{1} \equiv 0$ and $f(\Omega)$ is invariant
with respect to the inversion $\theta \rightarrow \pi -\theta $. 
The second one, $S_{2}$, is usually associated with the nematic order parameter and measures the (quadrupolar) orientational order of a nematic state along the direction of the magnetic field. 
Finally, $\Delta$ is nonzero only if there is a preferential direction of alignment within a plane {\em perpendicular} to the field.
In this case, the system possesses two mutually perpendicular nematic directors (one parallel to the field and one perpendicular to the field) and the nematic state is therefore {\em biaxial} ($BX$). If $\Delta$ is zero, the rods' projections are distributed randomly  in the azimuthal plane 
and the system is of uniaxial ($U$) symmetry.

The sign of $S_{2}$ is also important and allows one to distinguish two types of nematic order. 
First, if $0 <S_{2}\leq 1$ the particles are preferentially aligned along the field, corresponding to common {\em `polar'} nematic order.
Alternatively, if $-1/2\leq S_{2}<0$  the particles are mainly oriented in a plane perpendicular to the field direction, leading to  {\em `planar'} (or anti-nematic) order.  Note that the latter type  of nematic order only occurs  in systems
 subjected to disorientational external fields  and is not stable at zero-field conditions \cite{khokhlovsemenov}.

Once the ODF has been obtained, the thermodynamics and phase behavior of the system can be inferred from the osmotic pressure
and chemical potential. These are conveniently expressed in terms of the parameters 
$\sigma$, $\rho$ and $\eta$  [cf. Eqs. (\ref{rhosigma},\ref{eta})] via:
\begin{eqnarray}
\beta \tilde{\Pi} &=& c +  c^{2} \frac{\partial \phi g_{P}}{\partial \phi}  \left \{ \rho[f] + h \eta [f] \right \}  \nonumber \\
\beta \tilde{\mu} &=& \ln c + \sigma [f] + 2 c \left (  g_{P}+ \frac{\phi}{2}\frac{\partial  g_{P}}{\partial \phi} \right ) \left \{ \rho[f] + h \eta [f] \right \} \nonumber \\
&& +  \left \langle \beta U_{m}(\cos \theta) \right \rangle _{f}  
\label{coex}
\end{eqnarray}
At phase coexistence, these quantities must be equal in each of the coexisting phases.
Second order phase transitions from e.g.  uniaxial to  biaxial nematic symmetries can be localized by means of a first-order
bifurcation analysis, as discussed  in detail in  Ref. \cite{wensinkbiaxial}.

\section{Intrinsic rod properties}
\subsection{Magnetic properties}
The electronic and magnetic properties of the goethite rods have been extensively discussed in 
Ref. \cite{davidsongoethiet}. Here, we shall only briefly recall some of the  basic quantities we need as input for 
our calculations. First of all, the remanent magnetic dipole moment of the rods $\mu_{r}$ is estimated to be 
 $10^{3}$  $ \mu_{B} $  ($\mu_{B}$ is the Bohr magneton). 
The diamagnetic susceptibility  $ \Delta \chi $ at room temperature is 1.7 $10^{-3}$ and the average particle volume 
 $v_{0} =$ 5.6 $10^{-23}$ $m^{3}$. Using these numbers in Eq. (\ref{jeeka}) we obtain $J=2.28$ $T ^{-1}$ and $K=0.72$ $T ^{-2}$. 
These values need to be considered with some care because  the magnetic properties are size-dependent and the 
inherent size polydispersity of the system therefore leads to a considerable spread in $J$ and $K$.

To check whether these numbers are representative for the experimental goethite system we may trace  $S_{2}$
as a function of $B$ for a dilute (i.e. isotropic at zero-field) system and locate its zero-point. 
Beyond  this point, the nematic order parameter is negative which signifies a gradual change towards the planar-type nematic order
found at high field-strengths. Experimentally, the zero-point is located at 0.35 T whereas theoretically we find 1.809 T, irrespective of the density (at least in the para-nematic density range, as will become clear later).
The large discrepancy is attributed  to the incertainty in $K$ which depends sensitively on the particle size. 
We can achieve  much closer agreement by doubling this value such that $J = 2.28$ $T^{-1}$ and $K = 1.44$ $T^{-2}$.
This gives a zero-point at 0.552 T, in reasonable agreement with the experimental value. In the actual calculations
we shall fix $J$ at $2.28 $ $T^{-1}$ and vary $K$ to verify the sensitivity of the phase behavior 
with respect to a change of the (dia)magnetic properties.

\subsection{Electrostatic properties}
The double-layer potential around a charged colloid (with constant surface charge density) can be determined 
from the Poisson-Boltzmann (PB) equation which, in our case, must be solved for a cylindrical geometry.
At large distances, the electrostatic potential $\psi$ around a cylinder with diameter $D$ takes the Debye-H\"{u}ckel (DH) form \cite{stroobantslading}:
\begin{equation}
\beta \psi e =  \Gamma K_{0} (\kappa D/2)
\end{equation}
where $e$  is the elementary charge and $K_{0}$ a modified Bessel function. The proportionality constant $\Gamma$ 
depends upon the surface charge density $\sigma_{\text{el}}$ of the particles. 
For highly charged particles like goethite,  the linearized (DH) equation cannot be used to obtain $\Gamma$. 
Instead, the full (non-linear)  PB equation must be solved to determine its value. 
Approximate but accurate analytical solutions of the PB equation for a cylindrical geometry were obtained by Philip and Wooding \cite{philipwooding}  which allow for a 
straightforward calculation of $\Gamma$ by means of a simple iterative procedure. Once $\Gamma$ has been obtained the effective rod diameter
can easily be calculated from:
\begin{equation}
\frac{D_{\text{eff}}}{D}=1+(D\kappa )^{-1} \left \{ \ln \left [ \frac{\pi \Gamma ^{2} \exp[ - \kappa D]}{\kappa Q} \right ] + \gamma _{E} -(1/2) \right \}
\label{deff}
\end{equation}
where $\gamma_{E}$ is Euler's constant and $Q$ the Bjerrum length. Using  $\sigma_{\text{el}}\approx$ 0.2 $C/m^{2}$, 
$D \approx $ 15 $nm$ and ionic strength $I\approx$ 4 $\cdot10^{-2}$ $M$ we find $\kappa ^{-1} \approx 1.5$ $nm$ and $\Gamma \approx 1.0$ $\cdot10^{3}$ for the goethite rods. 
Eq. (\ref{deff}) then gives us $D_{\text{eff}}/D\approx 1.65$. For the twisting parameter we thus find  $h\approx  0.063 $. 
These results  indicate that the effect of twist is expected to be rather insignificant. The diameter ratio however is quite high so that the effective aspect ratio $L/D_{\text{eff}}$ of the spherocylinder is much smaller than that of the bare particle.

\section{Rods in directional external fields: General scenarios}

\begin{figure}
\includegraphics[width=8cm]{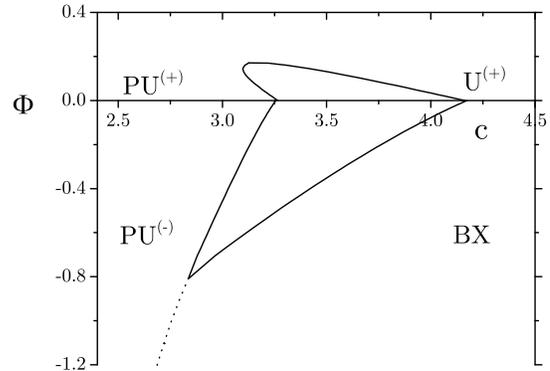}
\caption{Generalized para-nematic-nematic phase diagram for rods in external directional fields;
 $\Phi>0$ in case of an orientational field,  $\Phi<0$ for a disorientational field.
Paranematic-nematic binodals are given by solid lines, the dotted one represents a second-order phase transition from the para-uniaxial to the biaxial nematic state. }
\end{figure}
Before discussing the complex phase behavior of  the model goethite systems, we shall first present 
two simple general scenarios which occur if  rodlike particles are subjected to an external directional electro-magnetic field. 
The general phase diagram for this case has been presented in Fig. 1. This diagram is qualitatively similar to the one constructed by Khokhlov and Semenov in Ref. \cite{khokhlovsemenov} and has been recalculated here for infinitely thin hard 
rods ($L/D\rightarrow \infty$, $h$ =0) based on the free energy Eq. ({\ref{free}}).
For convenience, the external energy Eq. (\ref{magnenergy}) has been replaced by 
\begin{equation}
\beta U_{\text{ext}}(\cos \theta) = -\Phi {\cal P}_{2} (\cos \theta )
\end{equation}
in terms of a general field parameter $\Phi$  with $\theta$ the angle between the rod main axis and the field direction. 
 If $\Phi > 0$, the rods prefer to align along the field, and common `polar' nematic order occurs (indicated by ``(+)").
 In this case, the field is referred to as  having an `orientational' effect on the rods \cite{khokhlovsemenov}.  
 Note that due to the Boltzmann factor, $\exp[-\beta U_{m}]$, in Eq. (\ref{iter}), the isotropic state no longer exists at finite field-strengths. Instead, dilute systems now show weakly aligned para-nematic order, indicated by ``P". 
 Both nematic phases are of uniaxial symmetry and the first-order para-nematic-nematic coexistence region terminates
in a critical point above which the system changes from one state to  the other in a continuous fashion.
 As to magnetic fields, the present scenario may be observed for e.g. rods with a {\em positive} magnetic susceptibility, $\Delta \chi >0$, leading to an induced  moment along the main rod axis. The magnetic field then gives rise to liquid crystalline order of the {\em orientational quadrupolar} type  \cite{khokhlovsemenov}.
 Obviously, similar behavior is expected for rods with a permanent magnetic dipole moment along their main axis, like goethite (orientational dipolar field). In this case one refers to  {\em orientational dipolar}-type order.

In the opposite case  ($\Phi<0$)  the field has a `disorientational' effect and the rods preferentially orient in a plane
perpendicular to the field. Both para-nematic and  nematic phases are now of the planar, or anti-nematic type, indicated
by ``(-)". 
Moreover, in the concentrated nematic phase the rods apparently pack more favorably if they attain an additional direction of
alignment {\em within} the plane. The nematic phase thus has a biaxial symmetry.
For $\Phi<0$, the first-order para-nematic-nematic transition  terminates in a tricritical point \cite{vargakronome}. Beyond this point, the transition
from one state to  the other occurs by means of a second-order phase transition.
The present type of ordering may occur if the rods have a {\em negative} diamagnetic susceptibility anisometry so that
the induced magnetic moments are {\em perpendicular} to their main axes (disorientational quadrupolar order). This is the case for the induced moments of the goethite rods. Similar behavior has been found
recently for systems of colloidal gibbsite platelets in a magnetic field, where magnetic moments are induced along the short axis of the particle \cite{vdbeekSAXS}.

It is clear from the above that the goethite systems are expected to display characteristics from both scenarios.
This will become clear in the next section where we shall discuss some explicit phase diagrams for our model goethite systems.

\section{Phase diagrams for goethite}

\subsection{Quadrupolar scenario}
\begin{figure*}
\includegraphics[width=16cm]{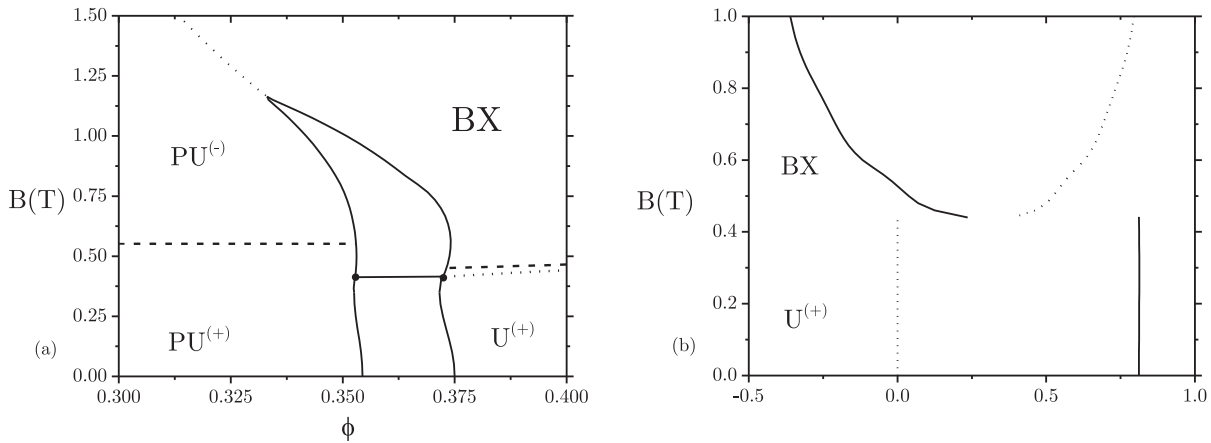}
\caption{(a) Phase diagram of `goethite' spherocylinders with $J=2.28$ $T^{-1}$, $K=1.44$ $T^{-2}$
and $L/D_{\text{eff}}=6.3$ ($L/D=10$).
Binodals are indicated by solid lines. The dashed line in the para-nematic region corresponds to $S_{2}=0$. 
The dotted curve in the para-nematic regime denotes a second-order phase transition
from the uniaxial $PU^{(-)}$ to the biaxial $BX$ nematic state, the one in the nematic regime 
represents a hard bifurcation  from $U^{(+)}$ to  $BX$. The dashed curve is the result from the Gaussian analysis (Appendix A). 
The corresponding behavior of the nematic order parameters $S_{2}$ (solid line) and $\Delta$ (dotted line)
 is shown in (b) for $\phi=0.40$. Note the  distinct jump at  $B=0.442$ $T$.}
\end{figure*}
Fig. 2a shows a phase diagram for `goethite' spherocylinders. 
This scenario is similar to the experimental situation, judging from the location of the
dotted line which marks the gradual change from polar to planar-type nematic order in the dilute regime, discussed in Sec. III-A.
We refer to this diagram as the `quadrupolar scenario' since the high-field region of the diagram 
is largely determined by the induced magnetic moments. Hence the appearance is similar to the disorientational quadrupolar scenario depicted in Fig. 1.
    
At low field strengths, the remanent moments dominate and the 
diagram is governed by the orientational effect of the field, i.e. the  para-nematic and nematic states are both uniaxial and the rods are strongly aligned along the field direction.
However, upon increasing the field strength, the degree of polar order will decrease since the induced moments (perpendicular
 to the main rod axis) become more pronounced. At some point, the uniaxial nematic state changes to a biaxial one and a first order (para-)uniaxial-biaxial nematic coexistence develops.  The coexistence region eventually narrows down towards a tricritical point, beyond which the $PU^{(-)}$-$BX$ transition becomes second-order \cite{khokhlovsemenov}. 
At very high field strengths, the induced moments will completely outweigh the remanent ones and force the rods to orient almost perfectly in a plane perpendicular to the field. The planar $PU-BX$ bifurcation then becomes reminiscent of a quasi-2D isotropic-nematic transition  \cite{vargakronome}.

Let us now focus on the field-induced transitions corresponding to the homogeneous (mono-phasic) systems, given by the horizontal curves in Fig. 2a.
In the dilute regime, the ODF changes continuously from polar-type (peaked around $\theta =0$ ) to planar-type (peaked around $\theta = \pi/2$ ) and the transition can be roughly localized from the condition $S_{2}=0$ (dashed curve). 
Note that the corresponding curve does not represent a phase transition, it merely localizes a gradual change of signature of the ODF. The curve is virtually independent of the packing fraction since the ODF in the dilute regime is mainly determined by the Boltzmann factor in Eq. (\ref{iter}).

In the concentrated regime, there is a transition from the  uniaxial to the biaxial nematic state corresponding to a hard bifurcation \cite{bauscryst}. This is evidenced by the behavior of the nematic order parameters in Fig. 2b, which display a distinct jump just above $B=0.4$ $T$. Note that this behavior is quite different from  a second-order phase transition (or soft bifurcation) where $\Delta$ rises {\em continuously} from zero, rather than jumping to a finite value. 
Physically, the order parameter jump can be associated with a sudden reorientation of the 
main nematic director from parallel to the field (at low  $B$) to perpendicular to the field (at high $B$). A similar phenomenon has been observed experimentally,
albeit at a somewhat lower applied field, $B \sim  0.2$  $T$ \cite{lemaireSAXS}.

It is important to note from Fig. 2b that the rods remain  sufficiently ordered along their main directors throughout the {\em entire} field range. By rotating its main nematic director perpendicular to the field the system is able to sustain the level of polar nematic order without changing to planar nematic order such as in the dilute regime.
 This particular property allows us to perform an asymptotic analysis of the free energy, valid for strongly ordered polar nematic states. In Appendix A,  we shall use Gaussian trial ODFs to approximately locate the transition and gain some insight into the underlying mechanism.
In the limiting case of perfectly aligned rods, valid for very dense nematic systems, the analysis yields a simple expression for the free energy difference between the $U^{(+)}$ and $BX$ phases at fixed concentration, which we will derive in an alternative way here.

If the spherocylinders are perfectly aligned, the $U^{(+)}$-phase (denoted by ``$\parallel $")
is represented by a collection of rods all parallel to the field. For the $BX$- phase (denoted by ``$\perp  $") we assume that all rods point along
a nematic director perpendicular to the field. At fixed concentration, the excluded-volume entropy $\rho$ (for $\gamma \equiv 0$) is identical in both states and the interactions therefore  do not contribute to the free energy difference.
 Conceptually, the system can be considered as an ideal ensemble  of dipoles, represented by spins.
 In the $\parallel $-state the magnetic energy  of a single spin $\beta U_{m}$ is then equal to $-JB + KB^{2}$ (parallel to the field) or  $JB+KB^{2}$ (anti-parallel). In the
 $\perp $-state all particles are perpendicular to the field and the magnetic energy is  invariant with respect to the direction of the spin,
hence $\beta U_{m} = -KB^{2}/2$ for all particles.  The free energy difference $\Delta {\cal F}_{\perp }-{\cal F}_{\parallel }$ is now easily calculated from the 
spin partition function and reads
\begin{equation}
\Delta {\cal F} = \ln [ \cosh JB] - \frac{3}{2}KB^{2} \label{spinfree}
\end{equation}
It is easily verified that $\Delta {\cal F} >0$ at low field strength
whereas $\Delta {\cal F} < 0 $ at high fields, indicating that perpendicular ($\perp $) order is indeed favored at high field strengths.
At the realignment transition  $\Delta {\cal F} $ is zero, and the corresponding field strength is found to be 
$B$ = 0.513 $T$ for the present case. This value represents an upper bound for the transition
and is not too far away from the numerical results.  The latter were obtained by numerically minimizing the free energy difference between the two states. The deviations are due to the orientational entropy
of the rods (neglected in the spin concept) for which we can partly account using the Gaussian approximation,
discussed in Appendix A.

\subsection{ Dipolar scenario}
 \begin{figure*}
\includegraphics[width=16cm]{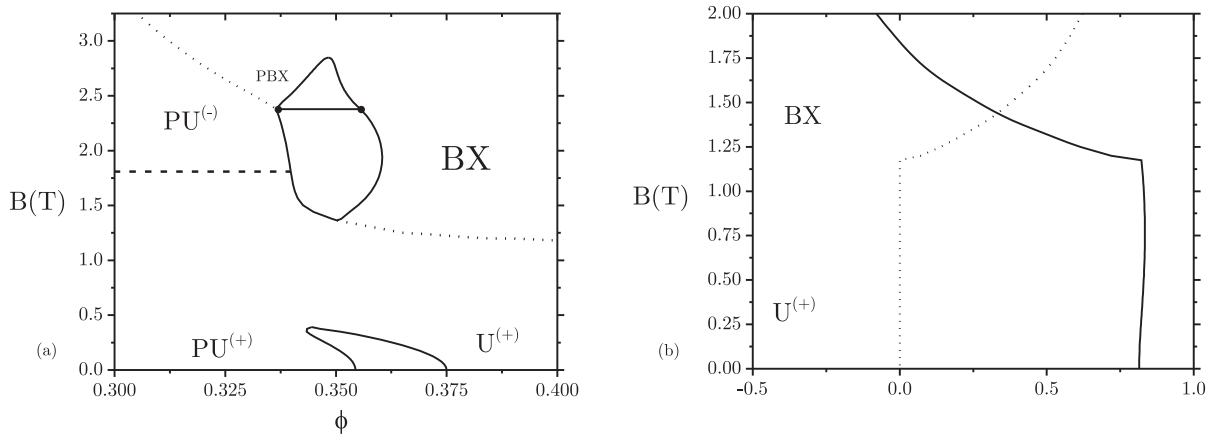}
\caption{(a) Phase diagram for $J=2.28$ $T^{-1}$ and $K=0.72$ $T^{-2}$. All dotted curves denote second-order phase transitions from the uniaxial to the biaxial nematic state. (b) Behavior of the nematic order parameters $S_{2}$ (solid line) and $\Delta$ (dotted line) as a function of $B$ for $\phi=0.40$.}
\end{figure*}
If we reduce the diamagnetic susceptibility anisometry (by lowering $K$) the diamagnetic effect becomes relatively unimportant at low fields.
The influence of the remanent dipole moments is then expected to govern the phase behavior 
in this regime. We see from Fig. 3a that the topology is indeed very similar to the orientational scenario of Fig. 1.
The para-nematic-nematic coexistence region terminates in a critical point above which the system changes gradually from one state to the other without any discontinuity or jump in the associated nematic order parameters.

At high field strength, the induced moments become dominant and give rise to a reopening of the phase gap beyond some 
critical $B$-value. Note that the nematic phase is now of biaxial symmetry, like in Fig. 2a.
A remarkable difference with the previous scenario however is that the para-nematic phase becomes biaxial as well 
at $B>2.38$ $T$. In this regime, a coexistence between two biaxial phases (a para-biaxial nematic ($PBX$) and
a biaxial nematic ($BX$) one) develops which eventually closes off at a critical or consolute point located at $B=$ 2.848 T.
Beyond this point the system gradually changes from one state to the other, similar to the para-nematic-nematic transition above the $PU^{(+)}-U^{(+)}$ critical point.
In fact, the para-biaxial-biaxial demixing region can be considered as the high-field analog of the $PU^{(+)}-U^{(+)}$ transition. Both involve a coexistence between phases of equal symmetry and the entropic mechanism underpinning the demixing is governed by a competition between orientational entropy (favoring the  weakly ordered para-nematic state) and packing entropy (favoring the nematic state).

An obvious consequence of reducing $K$ is that the transitions pertaining to the homogeneous  systems shift to much higher $B$-values, as we see in Fig. 3a.  
In the concentrated regime, the transition from the uniaxial to the biaxial state no longer corresponds to a hard bifurcation (or a realignment of the nematic director). Fig. 3b  shows that the nematic order parameters no longer display a real jump but merely a kink at the transition point indicating that we are dealing with a second-order phase transition, similar to the one occurring in the paranematic density regime.

\subsection{ Intermediate scenario}
\begin{figure}
\includegraphics[width=8.5cm]{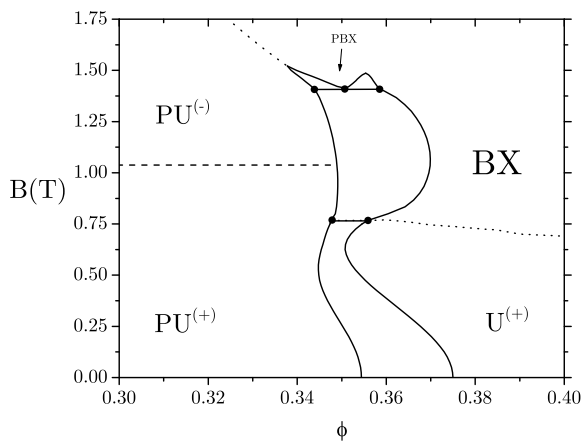}
\caption{ Phase diagram for $J=2.28$ $T^{-1}$ and $K=1.08$ $T^{-2}$. All dotted curves denote second-order phase transitions from the uniaxial to the biaxial nematic state. A triple line is located at $B=1.41$ $T$.}
\end{figure}
If we choose an intermediate value for $K$  we get an even richer phase diagram as can be seen from Fig. 4.
Close inspection reveals that the diagram contains features of both previous scenarios.
This is particularly notable at high field strengths, where we observe two
para-nematic-nematic regions  (involving  a $PU-BX$ and a $PBX-BX$ coexistence) reminiscent of the upper regions of Fig. 2a and Fig. 3a, respectively. Upon lowering $B$ the two regions meet 
at a triple line, indicating a triphasic coexistence between a uniaxial nematic phase and two biaxial nematic phases each with a different concentration. 

Marked reentrant and remixing effects are notable around $\phi = 0.36 $ where a sequence of phase coexistences may be expected upon increasing field strengths.  Similar to Fig. 3a, the transition  from the uniaxial to the biaxial  nematic state  in the  nematic density regime corresponds to a second-order phase transition, comparable to the one depicted in Fig. 3b.

The present scenario can be nicely  connected to the previous ones by focussing on the triple equilibrium.
If we increase $K$, the concentration of the para-biaxial phase (middle dot) is expected to move closer to 
that of the coexisting biaxial phase (right dot) so that the biaxial-biaxial region is pushed out of the diagram. At some $K$-value 
both concentrations meet at a critical end-point where the  $PBX$-$BX$ region has completely disappeared. From this point on the scenario will be similar to Fig. 2a.
If we decrease $K$, the opposite happens: the uniaxial-biaxial region is squeezed out at the benefit of the 
biaxial-biaxial region (Fig. 3a). Simultaneously, the lower $PU^{(+)}-U^{(+)}$ binodals
detach from the upper $PU^{(-)}-BX$ ones. The latter now constitute a separate coexistence region, enclosed by a lower tri-critical point and an upper consolute point. 

\subsection{Smectic order: frustration versus realignment}
\begin{figure*}
\includegraphics[width=16cm,height=10cm]{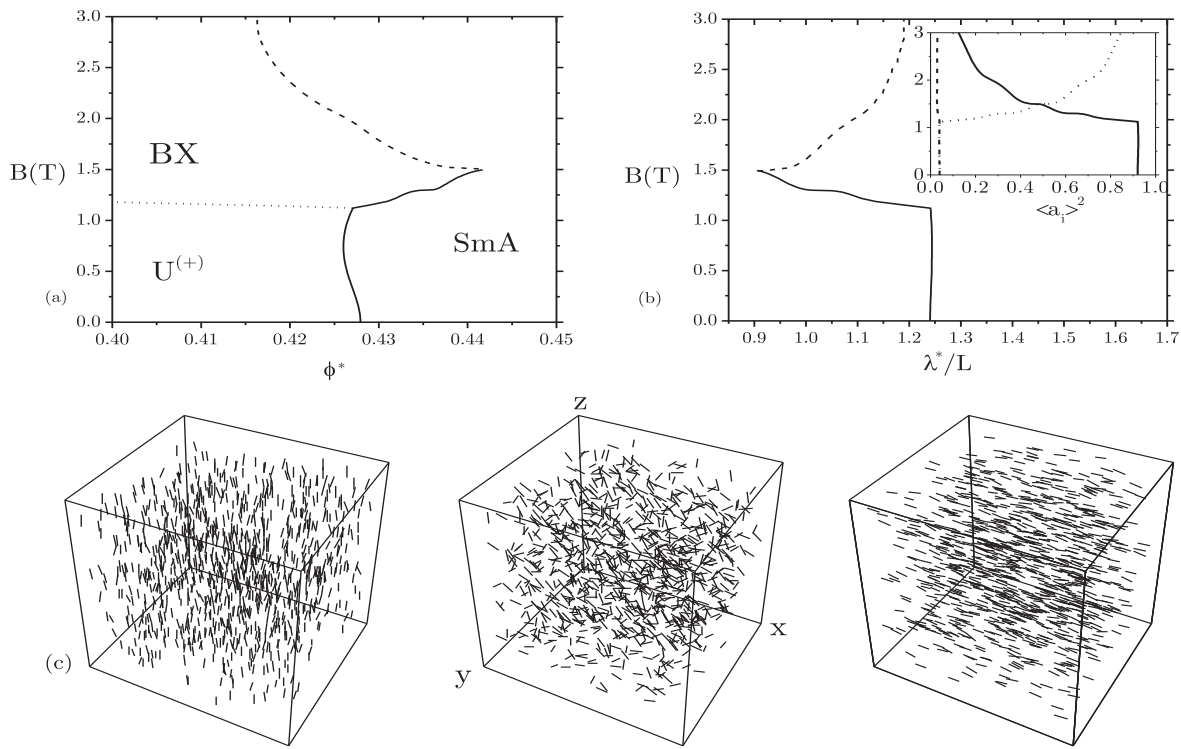}
\caption{(a) Nematic-smectic bifurcation for the `dipolar' scenario  ($J=2.28$ $T^{-1}$, $K=0.72$ $T^{-2}$).
The solid curve corresponds to smectic layering along the field (${\bf \hat{q}} \parallel {\bf \hat{B}}$), 
the dashed branch is associated with layering perpendicular to the field  (${\bf \hat{q}} \perp  {\bf \hat{B}}$).
The dotted line marks the second-order uniaxial-biaxial nematic transition.
The corresponding normalized layer spacings $\lambda _{\ast}/L$ are depicted in (b).
The inset shows the evolution of $ \left \langle a_{i}^{2} \right \rangle =\left \langle ({\bf \hat{u}}\cdot {\bf \hat{e}}_{i})^{2} \right \rangle $, the root-mean-square projection of the rod vector ${\bf \hat{u}}$ onto the reference axes ${\bf \hat{e}}_{i}$ ($i=x,y,z$), with $ \left \langle a_{x}^{2} \right \rangle $ (dotted), $ \left \langle a_{y}^{2} \right \rangle $ (dashed) and $ \left \langle a_{z}^{2} \right \rangle $ (solid). The magnetic field is directed along the $z$-axis. (c) Sketched snapshots representing the nematic phase
at the bifurcation density around $B=1.0$ $T$ (left) and $B=3.0$ $T$ (right). The middle one represents the quasi-isotropic structure
occurring around $B=1.5$ $T$.  
}
\end{figure*}
At packing fractions exceeding roughly 40 \% the nematic phase becomes unstable with respect to smectic order.
To investigate the implications of the magnetic field for the nematic-to-smectic transition we have 
performed a stability analysis of the nematic state with respect to small spatial density modulations pertaining to smectic order.
The bifurcation analysis is outlined in Appendix B. In Fig. 5 the results
are illustrated in terms of a bifurcation diagram for the dipolar scenario, corresponding to Fig. 3. 
At low field strengths, the nematic-smectic bifurcation is virtually the same as without external field, as we expect.
However, significant changes occur at the  transition to the biaxial state which, as we see from Fig. 3b,
 is associated with a continuous loss of polar nematic order (a decrease of $S_{2}$). 
As  the structure becomes more planar-like, the transition to the smectic state persists upon increasing
field strength but the associated layer spacing decreases rapidly and even becomes smaller than unity (Fig. 5b).
This can be explained from the fact that the rods are strongly tilted away from the field direction and therefore attain  an effective longitudinal dimension being smaller than the rod length $L$. 
Due to the biaxial signature of the nematic reference state, the structure of the smectic phase 
is also expected to be biaxial, although a full analysis of the (first-order) nematic to smectic transition is probably required to make conclusive statements about this.

At high  field strengths, the smectic phase eventually realigns in a similar fashion as the nematic phase.
In the realigned state, the smectic density modulations  propagate along the (main) nematic director {\em perpendicular} to the field. Similar to the parallel branch, pronounced planar-type order occurs (in this case in the $xz$-plane) upon lowering $B$ leading to a decrease of the layer spacing and an increase of the  bifurcation density. At very high field strengths, the nematic-smectic bifurcation becomes more or less analogous 
to the one at low fields.  Although the reoriented nematic structure remains (slightly) biaxial at high fields, the basic difference between the two will be merely a change of the laboratory frame.

The subtle evolution of the nematic structure upon variation of the field strength is perhaps more clearly reflected in the configurational snapshots and the inset of Fig 5b, showing the degree of order along the three axes of the laboratory frame as a function of $B$. Note that the field points along the $z$-axis of the frame. At low fields, the rods align parallel to field  so that $\left \langle a_{z}^{2} \right \rangle \gg \left \langle a_{x,y}^{2} \right \rangle$. In the reoriented nematic phase at very high fields, the rods are mostly pointing along the $x$-axis perpendicular to the field, hence $\left \langle a_{x}^{2} \right \rangle \gg \left \langle a_{y,z}^{2} \right \rangle$.  There is an intersection point where $\left \langle a_{z}^{2} \right \rangle  = \left \langle a_{x}^{2} \right \rangle$, indicating that all rod projections onto the $xz$-plane have equal probability.
The corresponding snapshot shows that the structure  appears more or less isotropic. However, it is not strictly isotropic  since the rod vectors only have a small $y$-component, $\left \langle a_{y}^{2} \right \rangle \ll \left \langle a_{x,z}^{2} \right \rangle $, indicating that they are strongly tilted away from  the $y-$axis.
Note that in the isotropic phase $\left \langle a_{x,y,z}^{2} \right \rangle  \equiv  1/3 $, irrespective of the choice of laboratory frame.

Going back to Fig 5, it is not too surprising 
that the two smectic branches meet at exactly the same state point. Both parallel and perpendicular smectic instabilities   are equally favorable because the degree of order is the same in both directions.
In fact, the quasi-isotropic orientation distribution in the $xz$-plane can be associated with a degeneracy of all smectic instability directions {\em within} the $xz-$ plane.

Although the present result may be mathematically sound, it remains rather awkward from a physical point of view.
In particular, one may question whether smectic order  will be  found at all  in such weakly aligned configurations.
We argue that the present results may be an artifact of the assumption that the ODF is constrained
at the bifurcation point (see Appendix B). The bifurcation densities presented in Fig. 4 may therefore deviate from those found from the exact eigenvalue equation \cite{roijtransverse}.
In the latter case,  subtle changes in the ODF are accounted for. These may, for instance, cause both branches to terminate  at a terminal point. This particular feature would be analogous to the presence of a terminal polydispersity for the nematic-to-smectic transition  found for parallel hard rods with length-polydispersity \cite{wensinkthesis}.
Therefore we anticipate that the applied magnetic field may give rise to a complete
frustration of smectic order in a small field interval around the intersection point.

For the intermediate scenario we expect similar features since the nematic structure realigns in an analogous fashion  compared to the one outlined in Fig. 5.  As to the quadrupolar scenario, different behavior is expected because the sudden reorientation of the nematic director may  also take place within the smectic phase. In that case, the intermediate quasi-isotropic configurations in Fig. 5 are less likely to occur and complete frustration of smectic order is probably less pronounced than in the other two cases.
It would be intriguing to  verify  the possible destruction of smectic order for the various cases by means of a computer simulation study. In this way  more conclusive insight could be generated about the implications of an applied magnetic field on the nematic-to-smectic transition.

Although in experiment, smectic order seems to be largely suppressed in favor of columnar order (irrespective of the applied field strength \cite{lemairecolumn}), smectic textures have very recently been observed  in between the nematic and the columnar phases \cite{vroegethies}.  It remains to be investigated what happens to these textures in an applied magnetic field.

\section{Conclusions}
Within the Onsager-Parsons theory we have investigated the stability of the various nematic states which may
appear in systems of goethite rods when subjected to an external magnetic field. 
In the present study, the goethite rods are represented by charged spherocylinders bearing a remanent magnetic moment
(leading to preferred dipolar order) and a {\em negative} diamagnetic susceptibility anisometry (leading 
to preferred planar, or quadrupolar, order). 
These mixed dipolar/quadrupolar properties give rise to intricate liquid crystalline phase behavior.
Depending on the relative contributions of the particles' remanent dipole moment and 
the negative magnetic susceptibility anisometry, several scenarios were constructed.
The quadrupolar scenario, in which the effect of the remanent moments is relatively small, is characterized 
by a sudden realignment of the nematic director at some critical field strength comparable to experimental findings.
We have shown that  the realignment phenomenon can be described appropriately using Gaussian trial ODFs.

Upon lowering the susceptibility anisometry, qualitatively different phase diagrams are found.
In the dipolar scenario, where diamagnetic effects only become manifest at high field strengths,
 two separate para-nematic-nematic coexistence regions are found at low and high fields, the latter 
involving two biaxial nematic phases. At intermediate susceptibilities, a triphasic coexistence
is found between a uniaxial para-nematic phase and two biaxial nematic ones. 
Preliminary results for the nematic-smectic transition reveal that a similar realignment transition may take place for 
the smectic state. Subtle phenomena occur in an interval around the realignment field strength  where 
smectic order may be suppressed completely.

In the present calculations we have not accounted for the bar-shaped geometry 
of the goethite particles. The inherent biaxial shape may have serious
 consequences for the phase diagrams presented here, in particular with respect to transitions to 
 the biaxial  nematic state.  We anticipate that biaxial order will be significantly 
 stabilized because the bar shape  makes them prone to  biaxial nematic order, even at zero-field.   
 Further complications, such as size polydispersity could also be addressed  by considering e.g. binary mixtures
 of two different-sized spherocylinders. Note that the size-dependency of the magnetic properties should then also be 
taken into account.  However, given the complex phase behavior of the monodisperse systems considered here one may
 question whether it is worthwhile to pursue in this direction.
  
 From an experimental point of view, a promising way to reconciliate the present model system with the experimental 
one could be to reduce both the intrinsic
 bar-shape and the size-spread of the colloids.  The first could be achieved  by coating the particles with a layer of silica which would render them more cylinder-like. 
 The coating procedure also opens up the possibility of introducing hard-particle interactions by applying 
a polymer-grafting of the silica-coated particles and redispersing them into a suitable apolar solvent. However we do not expect this modification to give significantly different  phase behavior since the electrostatic twist effect is of marginal importance and all phase diagrams presented here qualitatively apply  to `hard'  goethite rods as well. The second goal, reduction of  the polydispersity, can be reached using  various purification  and fractionation methods. In particular, reducing the particles' considerable length polydispersity would be desirable to enhance the stability of smectic order. Finally, a systematic variation of particle size is expected to influence the relative importance of permanent and induced magnetic moments, which would offer a means to address different theoretical scenarios.
These experimental topics are currently under investigation at the Van 't Hoff laboratory and significant progress has already been made.

\section*{Appendix A : Gaussian approximation}
For strongly aligned states the ODFs are significantly peaked around their main nematic directors
so that we may  adopt Gaussian trial functions \cite{OdijkLekkerkerker}  to describe the orientation distribution
in the various nematic states. To simplify matters here, we shall neglect the effect of twist which, being a minor effect anyway, is expected 
to be even less significant for the strongly aligned systems we consider. 

For the uniaxial state (denoted by ``$\parallel $") the following Gaussian trial ODF is proposed
\begin{equation}
f_{G}^{\parallel }(\theta) \propto  \exp \left [ - \frac{1}{2}  \alpha \theta ^{2} - \beta U_{m} (\cos \theta) \right ] \label{gaussalgemeen}
\end{equation}
in terms of the variational parameter $\alpha$.
Using  Eq. (\ref{magnenergy}) in the asymptotic limit ($\theta\ll 1$) and some rearranging gives:
\begin{equation}
f_{G}^{\parallel }(\theta) \simeq  Z \left\{ 
\begin{tabular}{lll}
$ \exp \left [- \frac{1}{2} \alpha_{(+) }  \theta ^{2} + JB \right ] $ & if & $%
0\leq \theta \leq \frac{\pi }{2}$ \\ 
&  &  \\ 
$ \exp \left [- \frac{1}{2} \alpha_{(-) }  (\pi - \theta) ^{2} - JB \right ]   $ 
& if & $\frac{\pi}{2}\leq \theta \leq \pi $
\end{tabular}
\right.  \label{0ODF}
\end{equation}
where $\alpha_{ (\pm)} = \alpha _{\parallel} \pm JB \gg 1$ and $\alpha _{\parallel} = \alpha -3KB^{2}$. The normalization factor $Z$ is a bit complicated and reads:
\begin{equation}
Z \sim  \frac{1}{4\pi}  \left \{ \sum _{\pm}  \exp[\pm JB] \left ( \frac{1}{2 \alpha _{(\pm)}} - \frac{1}{6 \alpha _{(\pm)}^{2}} \right )  \right \}^{-1} 
\end{equation}
With the Gaussian ODF all  necessary integrations can be performed
analytically by means of asymptotic expansions for large $\alpha_{\parallel}$.  After straightforward but lengthy calculations we get for the orientational entropy and magnetic energy, respectively
\begin{eqnarray}
\sigma_{\parallel}  &\sim &   \ln \alpha_{\parallel} -1 +JB \tanh JB - \ln [\cosh JB ] + \frac{2}{3}\alpha_{\parallel}^{-1}  \nonumber \\
&&  +  JB \alpha_{\parallel}^{-1} \left ( \tanh JB -JB \cosh ^{-2} JB \right ) \\
\left \langle  \beta U_{m} \right \rangle _{f}   & \equiv &   -JB S_{1} + KB^{2}S_{2}   \label{magns1s2}
\end{eqnarray}
with
\begin{eqnarray}
S_{1} &\sim &  \tanh JB \left ( 1- \alpha _{\parallel }^{-1} \right )  - JB \alpha_{\parallel}^{-1} \cosh ^{-2} JB  \nonumber \\ 
S_{2}  & \sim &    1-  3 \alpha_{\parallel}^{-1} 
\end{eqnarray}
containing all contributions up to ${\cal O}(\alpha_{\parallel}^{-2})$.
For the packing entropy $\rho_{\parallel }$ we use the leading order contribution of the asymptotic expansion performed
by Onsager \cite{onsager}:
\begin{equation}
\rho_{\parallel}  \sim 4/ \sqrt{\pi \alpha_{\parallel}} \label{rho}
\end{equation}
Inserting  these expressions into the free energy Eq. (\ref{free}) (with $h=0$) and minimizing with respect to $\alpha_{\parallel}$ yields:
\begin{equation}
\alpha_{\parallel} =\left \{ \frac{cg_{P}}{\sqrt{\pi}}+\sqrt{\left ( \frac{cg_{P}}{\sqrt{\pi}} \right )^{2} + \delta _{\parallel}} \right \}^{2}
\label{alphapara}
\end{equation}
with
\begin{equation}
 \delta _{\parallel} = \frac{2}{3}+2 JB \tanh JB - 3KB^{2}  \label{dpara}
\end{equation}
Consistency  requires that $cg_{P} \gg  \sqrt{\pi|\delta _{\parallel}|}$ in the regime   where $\delta _{\parallel} < 0$. 
Note that for $B=0$ (or $cg_{P}\gg 1$) the regular quadratic dependency $\alpha_{\parallel} \sim 4c^2g_{P}^{2}/\pi$ is recovered \cite{Vroege92}.

A similar treatment can now be given for  a  biaxial nematic state (denoted by ``$\perp $") which is strongly aligned along an in-plane director perpendicular
to the field. Introducing $ \psi $ as the angle between the remanent dipole moment along the particle's main axis and the director, the magnetic (Zeeman) energy of the dipole
is then proportional to ${\bf \hat{\mu}} \cdot {\bf \hat{B}} = \sin \psi \sin \varphi $ with $\varphi $ the azimuthal angle 
describing the orientation within the plane perpendicular to the director.
From  Eq. (\ref{gaussalgemeen}) the  ODF for the $\perp$-state can be written as:
\begin{equation}
f_{G}^{\perp }(\psi, \varphi ) \propto \exp  \left [  -\frac{\alpha _{\perp}}{2} \psi ^{2} + JB \psi \sin \varphi - \frac{3 KB^{2}}{2} \psi ^{2} \sin ^{2}  \varphi  \right ]
\label{gaussperp}
\end{equation}
for $\psi \ll 1$. The biaxial nature of the ODF is evidenced by the explicit $\varphi$-dependence.
Note that Eq. (\ref{gaussperp}) is  no longer a pure Gaussian  (also because of the Zeeman term linear in $\psi$) 
and the angular averages are not easily obtained for this case. For large $\alpha_{\perp} $ we can expand Eq. (\ref{gaussperp}) to quadratic order in $\psi$. Applying the normalization then yields:
\begin{eqnarray}
f_{G}^{\perp }(\psi, \varphi ) & \simeq  & \frac{\alpha_{\perp}}{4 \pi} \left ( 1 - \frac{3 {\cal K} B^{2}}{2 \alpha_{\perp}} - \frac{1}{3 \alpha _{\perp}}   \right )^{-1} \exp  \left [  -\frac{ \alpha_{\perp}}{2} \psi ^{2} \right ]   \nonumber \\
&& \times \left \{ 1  + JB \psi \sin \varphi - \frac{3 {\cal K} B^{2}}{2}  \psi ^{2} \sin ^{2}  \varphi  \right \}  
 \label{gaussperplin}
\end{eqnarray}
where ${\cal K} = K - J^{2}/3$. The orientational entropy in the $\perp $-state reads
\begin{equation}
\sigma_{\perp} \sim  \ln \alpha_{\perp} -1 +\frac{1}{\alpha_{\perp}} \left ( \frac{2}{3} + \frac{3}{2}KB^{2}  \right )
\end{equation}
and the nematic order parameters are given by 
\begin{eqnarray}
S_{1} &\sim & JB \alpha_{\perp}^{-1} \nonumber \\
S_{2} &\sim & (1/2) \left (3 \alpha_{\perp} ^{-1} -1 \right ) \label{orderparameters}
\end{eqnarray}
up to leading order in $\alpha_{\perp}$. The magnetic energy is then directly obtained from Eq. (\ref{magns1s2}).
For the $\perp $-state, however, it is more natural to define the nematic order parameters along the  nematic director ${\bf \hat{n}}$ rather than the field direction as in Eq. (\ref{orderparameters}). This gives: 
\begin{eqnarray}
S_{2}^{({\bf \hat{n}})} & \simeq & \left \langle 1-(3/2) \psi ^{2} \right \rangle_{f^{\perp}_G} \sim  1-  3\alpha_{\perp}^{-1}  \nonumber \\
\Delta ^{({\bf \hat{n}})} & \simeq  & \left \langle \psi^{2} \cos 2 \varphi \right \rangle_{f^{\perp}_G} \sim  3 {\cal K} B^{2} \alpha_{\perp}^{-2}
\end{eqnarray}
up to  leading order in $\alpha_{\perp}$. The latter result shows that the biaxial order parameter scales inverse quadratically
 with $\alpha_{\perp}$ and is therefore very small. Due to the inherent biaxial structure of the $\perp $-state, the packing entropy is difficult to access. Here we assume that the dominant contribution will be the result for the uniaxial state, $\rho_{\perp }\sim4/\sqrt{\pi\alpha_{\perp}}$
[cf. Eq. (\ref{rho})] and that  biaxiality is expected to give higher order corrections of minor importance.  
Minimizing the free energy for the $\perp$-state again leads to the form Eq. (\ref{alphapara}) but with
$\delta _{\parallel}$ replaced by:
\begin{equation}
\delta _{\perp} = \frac{2}{3} + 3 {\cal K} B^{2}
\label{alphaperp}
\end{equation}
Gathering results we arrive at the following free energy difference 
$\Delta {\cal F}={\cal F}_{\perp} - {\cal F}_{\parallel}$ (in units $k_{B}T$ per particle) between the two states:
\begin{eqnarray}
 \Delta {\cal F} && \sim   \ln[\cosh JB] - \frac{3}{2} KB^{2} +\ln \frac{\alpha_{\perp }}{\alpha_{\parallel}} \nonumber \\
&& + \frac{4cg_{P}(\phi)}{\sqrt{\pi}} \left [\alpha_{\perp}^{-1/2} - \alpha_{\parallel}^{-1/2}   \right ] 
+ \frac{ \delta _{\perp} }{\alpha_{\perp}}   - \frac{\delta _{\parallel}}{\alpha_{\parallel}}
\end{eqnarray}
containing all asymptotic contributions up to $ {\cal O }(\alpha ^{-2})$.
For dense systems, $cg_{P}({\phi}) \gg 1$ and $\alpha \gg 1$ so that the first two leading order terms survive
[cf. Eq. (\ref{spinfree})]. The $U^{(+)}-BX$ transition can be localized by solving $\Delta {\cal F} = 0$ 
with the aid of Eqs. (\ref{alphapara}), (\ref{dpara}) and (\ref{alphaperp}). The results of the Gaussian analysis are shown in Fig. 2a and are reasonably close to the numerical results.

\section*{Appendix B : Nematic to smectic-A transition}
To approximately locate phase transitions from the nematic to the smectic state we may apply a first-order
bifurcation analysis starting from the (para)nematic free energy  Eq. (\ref{free}) \cite{mulderparallelNS,roijtransverse}.
Also here, the effect of twist will be ignored, i.e. we set $h=0$. A limit of local stability for the nematic state with respect to infinitesimally small spatial density modulations is then associated with a divergence of the nematic structure function $S({\bf q})$. This leads to the condition
\begin{equation}
S^{-1}({\bf q}) \equiv 1-\phi g_{P}(\phi) \left \langle \left \langle \hat{f}_{M}({\bf q}; \Omega, \Omega ^{\prime})/v_{0} \right \rangle \right \rangle _{f} = 0
\label{bifur}
\end{equation}
where ${\bf q}$ is the wave vector pertaining to the direction of the density modulations. In principle, these could  represent smectic or columnar order. For (sphero)cylinders, the  nematic-columnar bifurcation  is strongly pre-empted by the nematic-smectic bifurcation \cite{mulderparallelNS} and we shall therefore not consider the possibility of columnar-type instabilities  for goethite rods. The shape of the rods  is enclosed in
$\hat{f}_{M}$, the cosine-transform of the excluded-volume body of two hard spherocylinders, which can be deduced analytically
[up to ${\cal O}(D^{3})$] from a straightforward but tedious geometric analysis \cite{vroijthesis}.

We have to realize that in our system smectic layering may occur either along the field direction (at small field strengths), so that ${\bf \hat{q}} = \{0,0,1 \}$ or perpendicular
to the field (at large field strengths), in which case ${\bf \hat{q}} = \{1,0,0\}$ or $\{0,1,0\}$. 
We stress that Eq. (\ref{bifur}) assumes a {\em fixed} ODF at the bifurcation. At the `exact' bifurcation point, changes 
in the ODF may also contribute to the loss of nematic stability. The corresponding eigenvalue equation
however is difficult to analyze numerically for  biaxial nematic reference states and we shall not consider it here.
The bifurcation to the smectic state is  given by the
wave vector $q^{\ast}=2\pi/\lambda^{\ast}$ (with $\lambda^{\ast}$ the characteristic layer spacing) which gives rise to the smallest physical solution $\phi^{\ast}$ of Eq. (\ref{bifur}).

For slender rods ($L/D\gg 1$) in strongly aligned configurations it is possible to derive simple asymptotic expressions for $\hat{f}_{M}$. Assuming 
small angular deviations from the nematic director, the cosine-transform can be written in the the following asymptotic
form (henceforth $D=D_{\text{eff}}$):
\begin{equation}
\hat{f}_{M}(\tilde{q};\Omega,\Omega^{\prime}) = -2L^{2}Dj_{0}^{2}(\tilde{q}/2)\sin \gamma(\Omega,\Omega^{\prime})-2\pi LD^{2}j_{0}(\tilde{q})
\end{equation}
with $ j_{0}(x)=x^{-1}\sin x$ a spherical Bessel function and  $\tilde{q}=qL$  the rescaled wave vector.
Using this result in Eq. (\ref{bifur}) and noting that the angular average of the sine is proportional to $\rho[f]$
we can write [with the aid of Eq. (\ref{rho})]:
\begin{equation}
S(\tilde{q})^{-1} = 1 +  j_{0}^{2}(\tilde{q}/2) \left( \frac{8 \phi g_{P} (\phi) } { \sqrt{\pi \alpha}} \frac{L}{D} \right )   +8 \phi g_{P}(\phi) j_{0}(\tilde{q}) \label{ast}
\end{equation}
where $c\sim \phi L/D$.  At zero-field, $\alpha \sim 4c^{2}g_{P}^{2}/\pi$ and the term between brackets reduces to $4$.
The corresponding asymptotic result for the nematic-smectic bifurcation is then found at $\phi^{\ast} =0.4037 $ with
corresponding layer spacing $\lambda^{\ast}/L=1.293$. The bifurcation density is in good agreement with the simulation result
$\phi^{\ast} =0.418$ \cite{Bolhuisintracing}.
  We stress  that the asymptotic result Eq. (\ref{ast}) can only be applied if the nematic order is strongly polar-like ($S_{2}$ close to unity). 
This is only the case  either at very low or very high field strengths where the nematic structure is not significantly disrupted by an imminent reorientation of the nematic director.

\acknowledgments
We want to thank Patrick Davidson for stimulating our interest in this problem.

\end{document}